\documentclass[aip,
 amsmath,amssymb,
 reprint,%
nofootinbib]{revtex4-2}

\usepackage{graphicx}
\usepackage{dcolumn}
\usepackage{bm}
\usepackage{xcolor}
\usepackage[utf8]{inputenc}
\usepackage[T1]{fontenc}
\usepackage{mathptmx}
\usepackage{etoolbox}
\usepackage{hyperref}
\hypersetup{
    colorlinks=true,
    linkcolor=blue,
    citecolor=magenta,
    urlcolor=blue
    }

\makeatletter
\def\@email#1#2{%
 \endgroup
 \patchcmd{\titleblock@produce}
  {\frontmatter@RRAPformat}
  {\frontmatter@RRAPformat{\produce@RRAP{*#1\href{mailto:#2}{#2}}}\frontmatter@RRAPformat}
  {}{}
}%
\makeatother
\begin{document}
\title{Thermometry with multilevel transmon probes}

\author{Antonio Mandarino}
\affiliation{International Centre for Theory of Quantum Technologies, University of Gda\'nsk, ul. prof. Marii Janion 4, 80-308 Gda\'nsk, Poland}
 \email{antonio.mandarino.work@gmail.com}
 \author{Matteo G. A. Paris}
 \author{Claudia Benedetti}
\affiliation{Dipartimento di Fisica "Aldo Pontremoli", Università degli Studi di Milano, I-20133 Milano, Italy}

\date{\today}

\begin{abstract}
Superconducting transmon systems are promising platforms for nanoscale thermometry due to their high sensitivity to environmental fluctuations. Their intrinsic anharmonicity, which is essential for qubit operations, gives rise to a non-equidistant energy spectrum that significantly affects the thermal populations and, consequently, the thermometric sensitivity.  
In this work, we  investigate the ultimate quantum limits of temperature estimation with a transmon beyond the two-level approximation. We compare the thermometric performance of three complementary models: the qubit, a harmonic oscillator and a weakly anharmonic Duffing oscillator, evaluating their corresponding quantum Fisher information (QFI) as a function of the temperature. 
We show that the multilevel anharmonic structure of the transmon affects its  thermometric precision. Indeed, including higher excited states enhances the maximum amount of   information that can be extracted about  the system temperature, compared to a qubit probe.
Furthermore, we address a fundamental limitation of the  standard quartic truncation, which yields a potential that is unbounded from below and supports only spurious  metastable states. By introducing bounded anharmonic models, namely a confined quartic potential and a sextic correction term, we assess the robustness of the thermometric precision  beyond the Duffing regime.  
Our results provide practical guidelines for transmon-based nanoscale thermometry  and clarify the role of the anharmonic multilevel spectrum in  quantum temperature estimation.

\end{abstract}

\maketitle

\section{Introduction}

In the past decades, we have witnessed an increasing experimental control over microscopic and mesoscopic systems, that has transformed quantum mechanics 
from a purely foundational theory into the operational framework of modern quantum technologies. 
In this trajectory, a particularly emblematic milestone was the discovery of macroscopic quantum mechanical tunnelling and energy quantisation in superconducting circuits,  definitively establishing that electrical circuits containing billions of electrons can behave as single coherent quantum objects \cite{clarke98,devoret13,Wendin_2017,Josephson2026}. 
This breakthrough provided the conceptual and experimental underpinnings of today's superconducting qubits, including the transmon architecture that now forms the backbone of many quantum processors and sensors. While the initial excitement around these systems was driven largely by their potential to implement quantum algorithms for tasks such as integer factoring, database search, and quantum simulation  \cite{nielsen2000quantum,Gaol2021PracticalGuide}, the same hardware platform has increasingly become a versatile playground for quantum-enhanced metrology and thermodynamics. In particular, the exquisite sensitivity of superconducting qubits to their electromagnetic and thermal environments, often regarded as an obstacle for fault-tolerant quantum computation, can be recast as a powerful resource for probing and controlling energy flows at the nanoscale \cite{QTT}.

Any realistic qubit is unavoidably coupled to thermal reservoirs, leading to relaxation and decoherence that degrade the coherent superpositions required for quantum information processing \cite{decoherence_thermalization2006}. 
In contemporary superconducting devices, these reservoirs may comprise engineered transmission lines, lossy resonators, spurious two-level defects, nonthermal quasiparticles, or driven nonequilibrium baths generated by control and readout circuitry \cite{Siddiqi2021,Muller_2019,catelani11,de_Graaf_2020}. 
Such structured environments can profoundly influence the qubit dynamics, giving rise to complex dissipative channels as well as to nontrivial steady states and transient regimes that carry  information about the underlying energy-exchange processes. From the perspective of quantum thermodynamics, a driven dissipative qubit coupled to its environment is no longer merely a noisy realization of an abstract two-level system; rather, it constitutes a prototypical thermodynamic nanoscale device capable of performing several tasks, ranging from managing heat currents to work extraction \cite{pekola2016finite, Goold_2016,ochoa2018quantum, Mandarino_2022, Gupta_2025, Li_2025}.
In this framework, the qubit-environment interaction can be harnessed for quantum thermometry,  enabling local probing of thermal fluctuations, effective temperatures, and energy-transfer processes in superconducting circuits and their surrounding electromagnetic environments \cite{Mehboudi_2019,Karimi2024}.

At present, the experimental platforms offered by superconducting circuit QED stands out as the most advanced candidate for near-term proof-of-principle thermotronic studies, owing to its combination of strong tunable nonlinearities, fine microwave control, engineered dissipation, and highly sensitive on-chip detection. Within this architecture, phenomena such as tunable photonic heat flow, thermal rectification, quantum-circuit refrigeration, and autonomous thermal-machine operation have already been realized, and the same setting has also been proposed for explicit implementations of quantum thermal transistors \cite{QTT, QTT_2, Gasparinetti_2025,ghosh2026quantumthermallogicgates}.
Superconducting transmon qubits are particularly well suited for such thermodynamic applications. Their anharmonic energy-level structure, reduced charge dispersion, and compatibility with circuit quantum electrodynamics enable high-fidelity control and readout over a broad range of frequencies and temperatures. 
Crucially for thermometry, the steady-state and transient populations of the lowest transmon levels encode information about the effective temperature of the composite environment coupled to the system.
Recent experiments \cite{Sultanov_2021, ThermoSC_exp} have shown that by applying carefully designed sequences of $\pi$-pulses to permute the populations of the lowest energy levels, 
one can reconstruct the occupation probabilities and thereby infer the temperature of the environment to which the transmon is thermalized. 
The attainable temperature resolution and signal-to-noise ratio are governed by the relaxation dynamics and coherence properties of the transmon, which ultimately limit the accuracy of the inferred temperature.
Another thermometry approach has been addressed in \cite{Zhu_2026}, in which the authors propose a coherence-mediated quantum thermometry scheme in a hybrid circuit-QED platform, where thermal fluctuations are transduced into a measurable phase shift on a coherent probe via coupling to a superconducting qubit. By exploiting this indirect readout mechanism, the protocol aims to preserve qubit coherence while enhancing sensitivity in the ultra-low-temperature regime, where conventional thermometric signals become extremely weak.

In this work, we employ the tools of local quantum estimation theory to explore the thermometric performance of a superconducting transmon beyond the two-level approximation \cite{paris09,brunelli11,Campbell_2018,razavian19, albarelli23}. While the standard qubit description is adequate for many quantum information tasks, thermometry requires the full treatment of the multi-level anharmonic structure. At finite temperatures, even when only a few  excited states are populated, the non-equidistant level spacing modifies the thermal response of the system and consequently the quantum Fisher information (QFI) that determines the ultimate temperature sensitivity.

The central question we address is how anharmonicity and the multilevel structure of the transmon affect its performance as a quantum thermometer. To answer this, we systematically analyze a hierarchy of effective descriptions,  ranging from the two-level qubit and the harmonic oscillator approximations to the  weakly anharmonic Duffing model derived from the Josephson potential. Within the formalism of local quantum  estimation theory, we evaluate the QFI for each model associated with temperature estimation, for experimentally relevant parameters in the transmon regime.

Our analysis shows that the multilevel anharmonic structure of the transmon has an impact on the thermometric performance. In particular, the Duffing oscillator beyond the qubit limit exhibits a larger QFI maximum at a finite optimal temperature, despite the harmonic model performing better at larger temperatures.  We further show that the thermometric precision can be tuned through  the Josephson energy, providing an experimentally accessible control parameter.

Finally, we assess the validity of the quartic approximation, which yields a potential that is unbounded from below,  by comparing it with bounded anharmonic models that incorporate higher order corrections to the Josephson potential. To ensure physically consistent predictions, we consider two regularized  models: a quartic potential confined by infinite barriers placed at its local maxima and a sextic potential obtained from the next-order term in the cosine expansion. Both approaches restores boundedness of the Hamiltonian while minimally perturbing the low-energy spectrum. This allows us to quantify the robustness of the thermometric precision and to identify regimes in which the Duffing description remains reliable. 

The paper is structured as follows. Section II reviews the derivation of the transmon Hamiltonian, the origin of anharmonicity, and the Duffing approximation. Section III introduces the mathematical framework of local quantum estimation theory, including the quantum Cramér–Rao bound and the expression of the QFI for thermal states. Section IV presents our results, comparing the thermometric performance of the qubit, harmonic, and Duffing models, and analyzing the dependence on truncation number and on the Josephson energy. Section V addresses the pathologies of the quartic approximation and introduces alternative physically consistent models, providing a rigorous benchmark for the thermometric predictions. Section VI concludes the paper with some final remarks.

\section{Anharmonicity in Superconducting Quantum Devices}
\label{sec:transmon}
The origin of harnessing superconducting platforms comes from experimental efforts to engineer macroscopic quantum tunneling in superconducting electrical circuits, which leads to energy-level quantization and macroscopic quantum coherence due to the absence of resistive losses \cite{Devoret1985, Martiniis1985}. 
Standard circuit elements, such as capacitors, inductors, and resistors were propelled by Josephson junctions, leading to systems whose dynamical description was given by the degrees of freedom corresponding to the collective charge and flux conjugate variables. 
The Josephson junction introduces a nonlinear and non-dissipative inductance, enabling the circuit to possess discrete, anharmonic energy levels suitable for quantum information processing. 
Several distinct types of superconducting qubits have been realized, including charge, flux, and phase qubits, \cite{ReviewAPR_qubits}, each defined by the relative magnitudes of the charging energy \(E_C\) and Josephson energy \(E_J\). Among these, the \emph{transmon qubit} has become the most widely used architecture due to its robustness to charge noise, and dispersion. The transmon consists of a single Josephson junction shunted by a large capacitance, operating in the regime \(E_J \gg E_C\). This design suppresses charge dispersion exponentially while retaining sufficient anharmonicity to isolate the lowest two quantum states, allowing coherent manipulation and scalability in circuit quantum electrodynamics (cQED) architectures \cite{tafuri2019fundamentals}.
%
Its quantum dynamics is described by the Hamiltonian
\begin{equation}
    \hat{H} = 4E_C\, \hat{n}^2 - E_J \,\cos \hat{\varphi},
    \label{eq:H_basic}
\end{equation}
where \(E_C = e^2/(2C)\) is the charging energy, with $e$ the electron charge and $C$ the junction capacitance, while \(E_J\) denotes the Josephson energy. The operator \(\hat{n}\) is the number of transferred Cooper pairs, while \(\hat{\varphi}\) is the superconducting phase difference across the junction. These conjugate variables satisfy the canonical commutation relation \([\hat{\varphi}, \hat{n}] = i. \)
  
In the \emph{transmon regime}, characterized by  \(E_J \gg E_C\), the phase fluctuations are small, allowing the cosine potential to be expanded in power series
resulting in the effective Hamiltonian 
\footnote{We neglect the constant contribution proportional to $E_J$. In the following, whenever additive constant terms  will be ignored since they do not affect the physical quantities considered here.}
\begin{equation}
    \hat{H} = 4E_C \hat{n}^2 + \frac{E_J}{2}\hat{\varphi}^2 - \frac{E_J}{4!}\hat{\varphi}^4 + \frac{E_J}{6!}\hat{\varphi}^6 + O (\hat{\varphi}^8), 
    \label{eq:H_expanded}
\end{equation}
The standard treatment of the transmon Hamiltonian retains the quadratic term, which describes an effective harmonic oscillator, together with the quartic contribution responsible for the weak anharmonicity needed to engineer the artificial atom. We first review the key steps leading to the conventional fourth-order truncation, neglecting the sextic term. The quadratic part of the Hamiltonian can be diagonalized by expressing the operators \(\hat{\varphi}\) and \(\hat{n}\) in terms of the bosonic creation and annihilation operators \(\hat{a}\) and \(\hat{a}^\dagger\) as follows:
\begin{equation}
    \hat{\varphi} = \varphi_{\mathrm{zpf}} (\hat{a} + \hat{a}^\dagger), 
    \quad
    \hat{n} = i\, n_{\mathrm{zpf}} (\hat{a}^\dagger - \hat{a}),
\end{equation}
where \(\varphi_{\mathrm{zpf}}= \left( \frac{2E_C}{E_J} \right)^{1/4}\) and \(n_{\mathrm{zpf}}= \left( \frac{E_J}{32E_C} \right)^{1/4}\) denote the zero-point fluctuation amplitudes, satisfying \(\varphi_{\mathrm{zpf}}\, n_{\mathrm{zpf}} = \frac12\).

Introducing the plasma frequency \(\hbar \omega_p = \sqrt{8 E_J E_C}\), the quadratic part of the Hamiltonian can be recast in the second-quantized form of a harmonic oscillator:
\begin{equation}
\label{H_ho}
    \hat{H}_0 = \hbar \omega_p \left( \hat{a}^\dagger \hat{a} + \frac{1}{2} \right).
\end{equation}
The quartic contribution arising from the Josephson potential is 
\begin{equation}
    \hat{H}_4 = -\frac{E_J}{24} \,\hat{\varphi}^4 
    = -\frac{E_J \varphi_{\mathrm{zpf}}^4}{24} (\hat{a} + \hat{a}^\dagger)^4 = -\frac{E_C}{12}\, (\hat{a} + \hat{a}^\dagger)^4.
\end{equation}
Keeping only the number-conserving terms in the quartic expansion yields
\begin{equation}
    \hat{H}_4 \approx -\frac{E_C}{2}\hat{a}^{\dagger 2}\hat{a}^2 - E_C \hat{a}^\dagger \hat{a} - \frac{E_C}{4}.
\end{equation}
Combining this quartic correction with the harmonic part gives
\begin{equation}
    \hat{H} = \hbar \omega_p \left( \hat{a}^\dagger \hat{a} + \frac{1}{2} \right)
    - E_C \hat{a}^\dagger \hat{a}
    - \frac{E_C}{2} \hat{a}^{\dagger 2}\hat{a}^2,
    \label{eq:H_eff}
\end{equation}
which describes a weakly anharmonic oscillator, commonly referred to as a Duffing oscillator.

Within this approximation, the Duffing energy levels are 
\begin{equation}
\label{eq:duffin_energies}
    E_n^D = \hbar \omega_p \left(n + \frac12\right) - \frac{E_C}{2}\, n (n + 1).
\end{equation}
As required for the implementation of an artificial atom, the transition frequencies between successive levels are nonuniform and decrease with increasing excitation number; explicitly,
\( \hbar \omega_{n,n+1} = E_{n+1}^D - E_{n}^D = \hbar \omega_p - E_C (n+1). \) 
A simple estimate shows that the number of physical levels is bounded by \( n < \left\lfloor 2\sqrt{\frac{2 E_J}{E_C}} - 1 \right\rfloor \), where \(\lfloor x \rfloor\) denotes the floor function (the greatest integer less than or equal to \(x\)).

The deviation from harmonic behavior—i.e., from equally spaced energy levels—is quantified by the anharmonicity parameter \(\alpha\), defined as the difference between two adjacent Bohr frequencies:
\begin{equation}
    \alpha = \hbar (\omega_{k, k+1} - \omega_{k-1,k}) = -E_C.
\end{equation}
Within the Duffing approximation, \(\alpha\) is constant across the spectrum. This negative anharmonicity enables selective driving of the \(|0\rangle \leftrightarrow |1\rangle\) transition while suppressing unwanted excitation of higher levels, thus allowing the transmon to operate effectively as a qubit.

At this stage, it is important to discuss the regime beyond the number-conserving approximation introduced in Eq.~\eqref{eq:H_eff}. The complete quartic term also contains non-number-conserving operators such as \(\hat{a}^2\), \(\hat{a}^{\dagger 2}\), and \(\hat{a}^{4}\), which couple Fock states differing by two or four excitations. Although these contributions are suppressed in the transmon regime \(E_J \gg E_C\), they induce small corrections to the level structure and enable weak multiphoton processes. The exact diagonalization of the truncated Hamiltonian \(\hat{H}_0 + \hat{H}_4\) can be recast as a well-known non-convergent problem, namely a quartic anharmonic oscillator with a negative quartic coupling. 

In Section~\ref{sec:beyondDuf}, we will discuss a theoretically consistent approach to treat perturbative effects beyond the number-conserving approximation.

\section{Quantum Thermometry}
\label{sec:quantum_thermometry}

Temperature estimation in the quantum regime can be naturally formulated as a parameter-estimation problem, where an unknown temperature \(T\) is encoded in a quantum state \(\rho_T\) and inferred from measurement outcomes. For \(N\) independent and identically prepared copies of the probe state, any unbiased estimator \(\tilde T\) satisfies the classical Cramér--Rao bound
\begin{equation}
\mathrm{Var}(\tilde T) \ge \frac{1}{N F_{\mathrm{cl}}(T)},
\end{equation}
where \(F_{\mathrm{cl}}(T)\) is the classical Fisher information associated with a given measurement \(\{\Pi_x\}\). For measurement outcomes  \(x\) occurring with probabilities
\begin{equation}
p(x|T) = \mathrm{Tr}\!\left[\rho_T \,\Pi_x\right],
\end{equation}
 the classical Fisher information is defined as
\begin{equation}
F_{\mathrm{cl}}(T)=\sum_x \frac{1}{p(x|T)}
\left(\frac{\partial p(x|T)}{\partial T}\right)^2.
\end{equation}
The ultimate precision allowed by quantum mechanics is obtained by maximizing the Fisher information  over all possible  quantum measurements. This leads to the quantum Fisher information (QFI) \(F_Q(T)\)
\begin{align}
    F_Q(T)=\mathrm{Tr}\!\left[\rho_T L_T^2\right]
\end{align} 
where $L_T$ is  the symmetric logarithmic derivative, implicitly defined through 
\begin{equation}
\frac{\partial \rho_T}{\partial T} = \frac{1}{2}\left(\rho_T L_T + L_T \rho_T\right).
\end{equation}
The quantum Cramér-Rao bound then reads
\begin{equation}
\mathrm{Var}(\tilde T) \ge \frac{1}{N F_Q(T)}.
\end{equation}
Since \(F_Q(T)\ge F_{\mathrm{cl}}(T)\) for any specific measurement strategy, the QFI sets the ultimate quantum limit to thermometric precision.

For thermal equilibrium states, the probe is typically described by the Gibbs state
\begin{equation}
\rho_T = \frac{e^{-H/(k_B T)}}{Z(T)},
\qquad
Z(T)=\mathrm{Tr}\!\left[e^{-H/(k_B T)}\right],
\end{equation}
where \(H\) is the system Hamiltonian, \(k_B\) is Boltzmann's constant and $Z(T)$ is the partition function. 
The QFI for a Gibbs state can be written as \cite{correa15}:
\begin{equation}
F_Q(T)=\frac{1}{k_B^2\, T^4} \,\Delta H^2
=\frac{1}{k_B^2 T^2}\, C_V,
\end{equation}
where \(
\Delta H^2 =\langle H^2\rangle-\langle H\rangle^2,\) is the energy variance and 
and \(C_V\) is the heat capacity defined as: \( C_V = \frac{\partial \langle H\rangle}{\partial T} = \frac{1}{k_B T^2}\Delta H^2. \)
This relation shows that thermometric sensitivity is directly governed by thermal energy fluctuations: the larger the heat capacity, the larger the Fisher information and the smaller the achievable estimation error.
Remarkably, energy measurement is optimal for equilibrium thermometry. Since the Gibbs state is diagonal in the energy eigenbasis \(\{|n\rangle\}\), 
\begin{equation}
\rho_T = \sum_n p_n(T)\,|n\rangle\langle n|,
\qquad
p_n(T)=\frac{e^{-E_n/(k_B T)}}{Z(T)},
\end{equation}
the projective measurement onto the Hamiltonian eigenstates \(E_n = |n\rangle\langle n|\) saturates the QFI, thus extracting all the temperature information contained in the state. The corresponding classical Fisher information 
\begin{equation}
F_{\mathrm{E}}(T)=\sum_n \frac{1}{p_n(T)}
\left(\frac{\partial p_n(T)}{\partial T}\right)^2
=F_Q(T)\,,
\end{equation}
equals the QFI. Hence, for equilibrium probes the quantum Cramér--Rao bound can be in principle saturated  by energy-resolved measurements.
\newline
The performance of a thermometric protocol may be quantified through the  signal-to-noise ratio
\begin{align}
R(T)=\cfrac{T^2}{\rm{Var}(\tilde{T})},
\end{align}
which measures the  relative precision achievable by an unbiased estimator $\tilde{T}$. 
Owing to the quantum Cramér-Rao bound, this quantity is upper bounded by the quantum signal-to-noise-ratio (QSNR)\cite{paris09},  $R(T)\leq R_Q(T)$, defined as:
\begin{align}
R_Q(T)=T^2\,F_Q(T).
\end{align}
Since the thermometric performance of a quantum probe is  determined by the temperature dependence of the QFI, at low-temperature the spectral gap becomes a crucial factor. When the thermal energy  \(k_B T\) is much smaller than the characteristic excitation energy , the population of excited states is exponentially suppressed reducing the probe's sensitivity to temperature variations and leading to a rapid decrease of the QFI.  Conversely, near critical points the enhanced thermal response of the system, often associated with heat-capacity peaks, can substantially increase the QFI, making quantum critical systems attractive candidates for precision thermometry \cite{salvatori_2014, De_Pasquale_2018, Mehboudi_2019,LandauB}.
From now on, we will be working in a system of natural units imposing $\hbar=k_B= 1.$

\section{Transmon as Thermomether }
Transmon qubits are often modeled as multi-level quantum Duffing oscillators \cite{2qubit_coupler}, an effective description that captures their anharmonic potential and enables precise control of their energy-level structure for quantum sensing applications. 
In the transmon regime with $E_J/E_C = 50$ (corresponding to $E_C \approx 0.2,\text{GHz}$ and $E_J \approx 10,\text{GHz}$), the system exhibits an almost flat energy spectrum as a function of offset charge, making it highly suitable for temperature estimation via quantum Fisher information. This flat spectrum and the transmon's robust coherence properties allow it to serve as a sensitive quantum thermometer \cite{Wendin_2017}.
For sake of simplicity, we further express all Hamiltonians in units of the plasma frequency, $ \omega_p$.
 \newline
 %
\begin{figure}[t]
    \centering
\includegraphics[width=0.97\columnwidth]{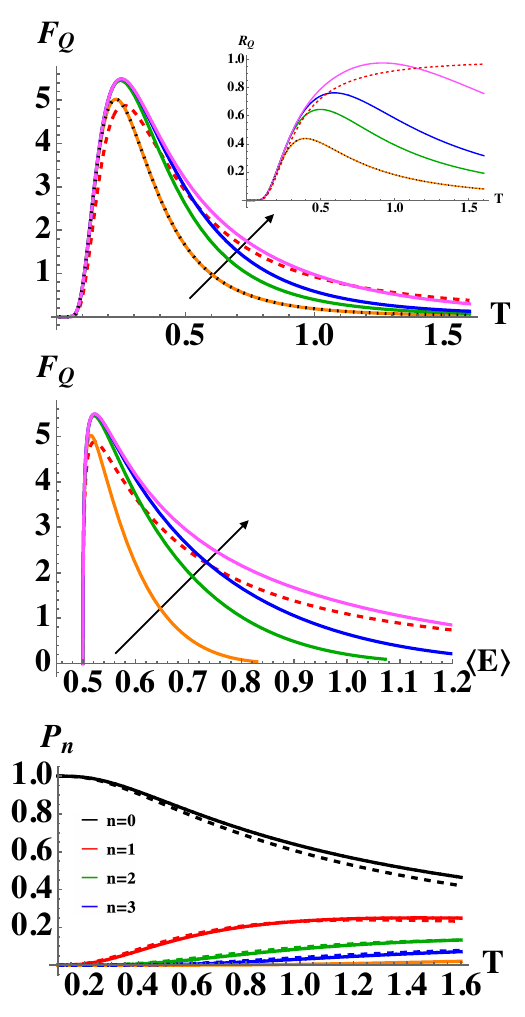}
\caption{Top: QFI as a function of temperature for different systems: the qubit (dotted black), the quantum harmonic oscillator (dashed red), and the Duffing oscillator with two (orange), three (green), four (blue) and eight (magenta) truncated levels. Inset: Quantum signal-to-noise ratio as a function of temperature, for the same systems. Center: QFI as a function of the mean energy, for the same systems as in the upper plot. The thin arrow is a guide for the eye for increasing dimensions of the Duffing model. Bottom: Thermal occupation probability $P_n$ of the energy levels for the harmonic oscillator (solid lines) and the Duffing oscillator (dashed lines).}
\label{fig_qfiD} 
\end{figure}
%

In the following, we consider three effective descriptions of the transmon system. The most accurate is provided by the Duffing-oscillator model, which incorporates the leading anharmonic correction to the Josephson potential. Its normalized spectrum, expressed in units of $ \omega_p$, is 
 $E^D_n= n + \cfrac12 - \cfrac12 \sqrt{\dfrac{Ec}{8 Ej}}\, n \,(n + 1)$.
 Neglecting the anharmonic contribution yields the harmonic-oscillator limit, described by Eq. \eqref{H_ho}, with equally spaced energy levels. Finally, restricting the dynamics to the two lowest levels leads to the qubit approximation, with effective Hamiltonian
 $$\hat{H}_2= {\cfrac12} \left(1 -  \sqrt{\dfrac{ E_C}{8E_J}}\right) \hat \sigma_z=\frac{\Delta}{2}\, \hat \sigma_z.$$
 The qubit energy splitting  $\Delta=E^D_1-E^D_0=\left(1 -  \sqrt{{ E_C}/({8E_J})}\right)$ coincides with the transition energy between the two lowest levels of  the Duffing spectrum.\\
 The QFI for the qubit and the quantum harmonic oscillator  can be obtained analytically. For the harmonic oscillator at frequency $\omega_p=1$, the quantum Fisher information is:
\begin{align}
F_Q^{\rm{(HO)}}=\left(\cfrac{1}{2T}\right)^2 \sinh^{-2}\left(\cfrac{1}{2T}\right),
\end{align}
while for a qubit with splitting energy $\Delta$,
 \begin{align}
     F_Q^{(2)}(T)=\left(\cfrac{\Delta}{2T}\right)^2 \cosh^{-2}\left(\cfrac{\Delta}{2T}\right).
 \end{align}
 $F^{(2)}_Q(T)$  reproduces the QFI of the Duffing oscillator when only its two lowest energy levels are retained.
 
In Fig. \ref{fig_qfiD}, we report the  QFI as a function of  temperature (top panel)  for the  the qubit, the quantum harmonic oscillator and the Duffing oscillator with different truncation dimensions. As the number of considered  Duffing levels is increased, the QFI becomes larger over the temperature range relevant for transmon systems, indicating an enhanced sensitivity to temperature. 
At higher temperatures, the quantum harmonic oscillator provides the largest QFI, as expected from its unbounded spectrum and the increasing contribution of highly excited states.
This behaviour is further confirmed  by  the QSNR shown in the inset of the top panel. In particular,  the effect of the Hilbert-space truncation becomes more pronounced in the QSNR, with larger truncation dimensions leading to improved thermometric precision over a wider temperature range.
For completeness, the central panel shows the  QFI as a function of the mean energy of the probe, $\langle E \rangle=\sum P_n(T)\,E_n$, where $E_n$ denote the energy levels of the system and $P_n(T)=e^{-E_n/T}/Z$ are the corresponding occupation probabilities.
We also observe that the maximum QFI rapidly converges as the number of retained Duffing levels increases, indicating that higher excited states provide only a negligible contribution to the optimal sensitivity. 
This is consistent with the occupation probabilities $P_n(T)$ of the Duffing levels,  shown in the bottom panel. Within the temperature range of interest, the population is predominantly concentrated in the lowest states, while higher-energy levels become progressively populated only as the temperature increases. Consequently, the contribution of additional levels to the QFI remains limited, explaining the observed saturation of its maximum value.
\begin{figure}[t]
    \centering
\includegraphics[width=0.98\columnwidth]{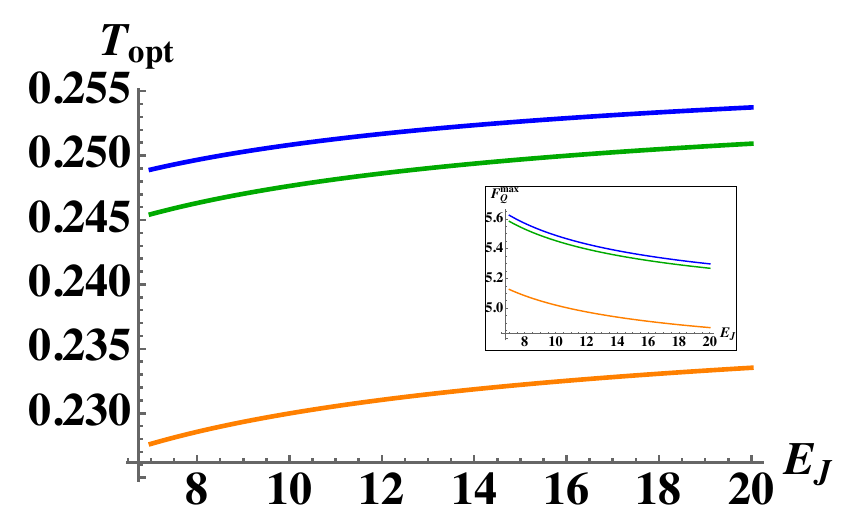}
\caption{Optimal temperature associated with the maximum quantum Fisher information as a function of the Josephson energy $E_J$, for a fixed charge energy $E_C=0.2$. Results are shown for the Duffing model with different truncation dimensions. The considered values of $E_J$ correspond to the range $E_J/E_C \in [35,100]$. The inset reports the corresponding maximum QFI.}
\label{fig_ratio}
\end{figure}
Nonetheless, a description beyond the qubit approximation provides a more accurate probe for quantum thermometry. The inclusion of higher energy levels results in an enhanced temperature sensitivity compared to the two-level model.

To complete our analysis, in Fig. \ref{fig_ratio} we investigate how the maximum of the QFI and the corresponding optimal temperature depend on  the Josephson energy $E_J$, while keeping the charge energy fixed at $E_c=0.2$. 
As $E_j$ increases, the optimal temperature shifts  slightly toward higher values, whereas the corresponding QFI  decreases. 
Consistent with the results discussed above, increasing the dimension of the truncated Duffing  space leads to improved thermometric performance compared with the two-level approximation, yielding a larger QFI over the relevant parameter range.

{\section{Thermometry  beyond the Duffing model}}
\label{sec:beyondDuf}
As discussed in Section \ref{sec:transmon}, the Duffing oscillator provides an accurate and widely used effective description of the low-energy transmon spectrum, thus becoming one of the most viable solutions for the practical treatment of a quantum computational unit. 
Nevertheless, the expansion of the Josephson potential in Eq. \eqref{eq:H_basic} presents several formal limitations that should be carefully assessed when investigating thermometric protocols based on multilevel transmons.
Let us start analyzing the Hamiltonian as reported in Eq. \eqref{eq:H_eff}. 
When the expansion of the cosine potential is truncated at fourth order and the perturbative regime around the minimum is considered, the resulting potential becomes unbounded from below at large displacements: the potential exhibits a local minimum at the origin, bordered by two symmetric local maxima, beyond which the potential diverges to negative infinity, making the system unbounded and unstable.
This pathology is solely an artifact of the quartic truncation, since the full cosine potential remains bounded from below and supports stable bound states.
nevertheless,  the low-energy states of the quartic potential can 
still be interpreted in terms of metastable resonances-quasi-stationary states whose lifetimes are determined by the tunneling through the finite barrier. In this regime, perturbative expansion around the harmonic oscillator diverges rapidly and non-perturbative methods, such as the instanton calculus, are required to obtain reliable spectral estimates \cite{zinn2021quantum}.

To assess the robustness of the thermometric results beyond the Duffing approximation, we consider two anharmonic models that remain well defined at all energies. 
Our theoretical analysis is formulated by considering a unitary mass harmonic oscillator described by the standard Hamiltonian $\hat H = \frac{\hat p^2}{2 m } + \frac{\omega^2 \hat x^2}{2}$ combined with two anharmonic potentials that are not perturbatively treated. 
In the following, the canonical variables $\hat{x}$ and $\hat{p}$ are identified with the flux $\hat{\varphi}$ and charge density $\hat{n}$, respectively, as discussed in Section \ref{sec:transmon}. Following Refs.~\cite{del_Valle_2023,Brevi2024,Brevi2024_tut}, we complement the results of section \ref{sec:quantum_thermometry}, with two practical but theoretically tailored strategies:
(i) engineer a quartic potential  with infinite potential barriers located at the positions of the local maxima $\hat V_4 (\hat \varphi) = - \frac{\lambda}{4!} \varphi^4= - \frac {1}{4!} \sqrt{\frac{ E_C}{2 E_J}}\varphi^4$ 
    and perform the numerical diagonalization within the interval bounded by the two maxima in $\pm \sqrt{\frac{6}{\lambda}} $; \\
(ii) consider also the sextic term appearing in Eq.\eqref{eq:H_expanded}, such that 
    $\hat V_6 (\hat \varphi) = - \frac{\lambda}{4!} \hat \varphi^4 + \frac{\mu}{4!} \hat \varphi^6  = - \frac {1}{4!} \sqrt{\frac{ E_C}{2 E_J}} \hat \varphi^4 + \frac {1}{6!} \frac{ E_C}{ E_J} \hat \varphi^6 $. 
Such a term only weakly affects to the energy shifts but ensures the boundness of the Hamiltonian and therefore the convergence to physical states of the anharmonic oscillator. The last equalities in  both the definitions of $V_4 (\hat \varphi)$ and $V_6 (\hat \varphi)$ make it explicit their connection with the transmon expansion in Eq.\eqref{eq:H_expanded}, after renormalization with respect to the plasma frequency $ \omega_p.$ 
To explore parameter regimes relevant to currently available transmon devices, we set $\lambda =0.1$ and $\mu=0.02$, corresponding to $\sqrt{\frac{E_C}{8 EJ}} = 0.05.$

\begin{figure}[t]
    \centering
\includegraphics[width=0.98\columnwidth]{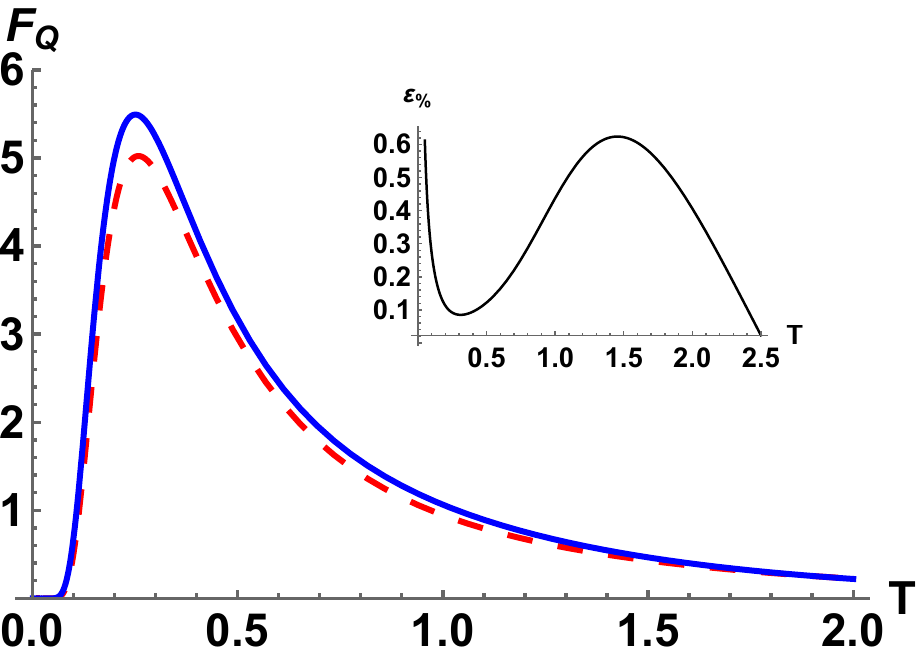}
\caption{Comparison QFI as a function of $T$ for Duffing (solid blue) and correction up to sextic (red dashed). In the inset, the relative percentage error computed as $ \frac{ F_Q[V_4]-F_Q[V_6]}{F_Q[V_6]} * 100$ All the curves are computed taking into account 12 energy levels.}
\label{fig_qfi6}
\end{figure}

\begin{figure}[t]
    \centering
\includegraphics[width=0.98\columnwidth]{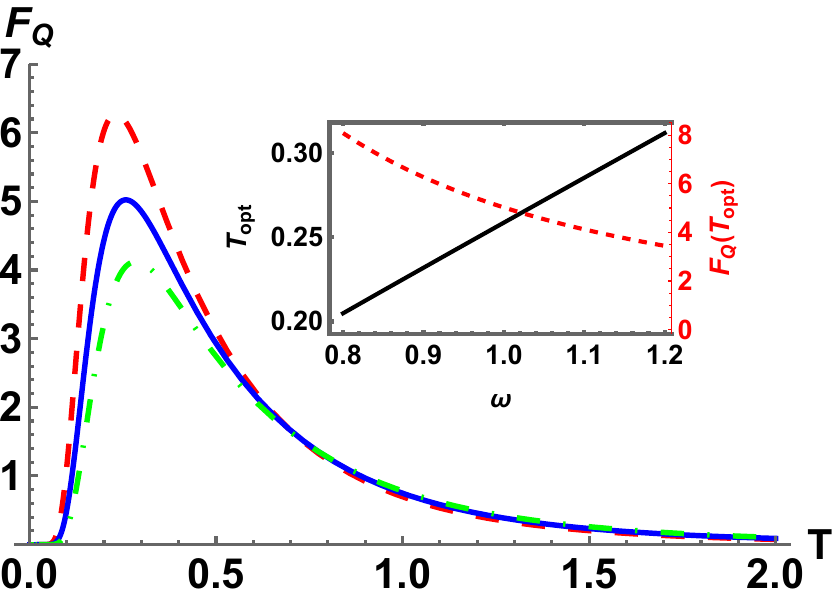}
\caption{Comparison QFI as a function of $T$ for various frequency $\omega$ of the anharmonic oscillator at fixed  $\omega=1$ (solid blue line), $\omega=0.9$ (dashed red line) and $\omega=1.1$ (dot-dashed green line). In the inset, the optimal temperature (left vertical axis) and the maximal Fisher information achieved at the optimal temperature (right vertical axis). For all curves, we have fixed $\lambda = 0.1$, and five energy levels. }
\label{fig_qfi_freq}
\end{figure}

The main results are reported in Fig. \ref{fig_qfi6}, where we compare the thermometric performance of the Duffing approximation with that obtained from the two bounded anharmonic models introduced above. 
The first noticeable difference concerns the energy of the lowest transition, which is slightly underestimated within the Duffing approximation, as it is $\omega_{01}^D = 0.95$ 
while the numerical results of our refined methods  yields $\omega_{01}[\hat V_6] \approx 0.9875 $. As a consequence, the corresponding quantum Fisher information is mildly overestimated (see also the discussion on Fig. \ref{fig_qfi_freq}).
It is worth noting that, for our thermometric goals, provided that the number of energy levels remains within the validity regime of  Duffing approximation \footnote{for the energy scales considered in the present manuscript, no more than 15 levels are involved}, the two anharmonic description  $V_4(\hat \varphi)$ and $V_6(\hat \varphi)$
yields essentially the same results. 
This observation is quantified in the inset of Fig. \ref{fig_qfi6}, which reports the relative error between the QFI  obtained from the two models. 
In the working temperature range of the superconducting circuit, this discrepancy varies between $0.1-0.6 \%$, demonstrating that both theoretical methods 
are suitable for practical purposes. At the same time, the sextic potential guarantees the convergence of the numerical schemes considered on every energy scale. 
Lastly, in Fig.\ref{fig_qfi_freq} we assess the scaling of the thermometric performances as a function of the harmonic frequency $\omega$ at a fixed level of anharmonicity. From an engineering point of view, this analysis has a twofold significance. On the one hand, it illustrates how different implementations of the effective LC circuit affect the optimal temperature of the estimation protocol. On the other hand,  it can be interpreted as a consequence of imperfect calibration of the superconducting device. Indeed, variations of the circuit parameters modify the optimal estimation temperature and the achievable precision, but do not significantly alter the overall thermometric performance of the multilevel transmon.
In this way, we obtain a rigorous controlled formulation to assess the robustness of multilevel transmon thermometry beyond the Duffing approximation. 
From an experimental perspective, the Josephson energy \(E_J\) constitutes the most accessible tuning parameter, and it is easier to manipulate with respect to the charging energy \(E_C\). In fact, \(E_J\) can be tuned in situ, e.g., using a SQUID loop or magnetic flux, and some studies explicitly show that magnetic fields modulate \(E_J\) while \(E_C\) remains essentially unaffected~\cite{transmon_original, ReviewAPR_qubits}. \(E_C\) is determined by the device total capacitance and is essentially fixed at fabrication, requiring geometric redesign rather than a control knob to change~\cite{2qubit_coupler}. 
Consequently, \(E_J\) serves as the primary experimental handle for frequency and anharmonicity adjustment, while \(E_C\) is mainly a design parameter chosen to achieve the large \(E_J/E_C\) ratio characteristic of the transmon regime \cite{Hutchings_2017, Wang_12lev}. 

\section{Conclusion}
Superconducting circuits emerged as one of the leading platforms to implement artificial atoms and two-level systems. 
Current quantum processors based on superconducting qubits outperform in number the competing platforms, and they have proven their versatility for the implementation of many quantum technological applications ranging from digital quantum simulation \cite{Grossi_2023, Kiss_2025} to quantum engines \cite{Gasparinetti_2025} and other quantum thermodynamical devices \cite{Josephson_QHE_2016}. 

The thermometric protocol discussed in this work is compatible with currently available transmon technology. State-of-the-art flux-tunable transmons routinely operate in the regime \(E_J/E_C \gtrsim 30\text{--}100\), with representative values \(E_C \sim 0.1\text{--}0.4\,\mathrm{GHz}\) and \(E_J \sim 5\text{--}20\,\mathrm{GHz}\), depending on the specific circuit design. 
In this parameter range, the transmon remains sufficiently anharmonic to permit selective addressing of the lowest transition (therefore working as a proper qubit for all logical implications), 
while its low-lying excited states remain thermally accessible in a controlled manner. 
The feasibility to control transitions between different energy levels and a non-negligible partial thermal occupation of higher excited states is what makes the device well-suited for equilibrium thermometry, 
since the temperature to be estimated is then encoded in the populations of a few lowest-energy levels, but beyond the qubit truncation.

The temperature range in which the transmon operates as an effective thermometer is determined by the interplay between the energy level spacing and the thermal population of the first excited states. 
For the parameter regime considered in this work, the optimal thermometric sensitivity is expected in th  few-to-hundred-millikelvin range, roughly \(T \sim 50\text{--}500\,\mathrm{mK}\), where the thermal energy becomes comparable to the lowest transition frequencies of the device. This operating regime is already accessible in dilution-refrigerator experiments, and it overlaps with the conditions for superconducting qubits used in recent thermometry demonstrations. 
At significantly lower temperatures, thermal occupation of excited states becomes exponentially suppressed, the QFI decreases rapidly and  the amount of temperature information encoded in the probe is reduced. Conversely, at higher temperatures, an increasingly large number of energy levels contributes to the thermal state, making a multilevel description essential for accurate thermometry.

Our results may be summarized as follows. 
First, we have analyzed thermometry with multilevel transmon systems beyond both the qubit and the harmonic approximation. We have shown that restricting the probe to  its two lowest levels does not fully exploit the thermometric resources offered by the transmon spectrum leading to  reduce values of the QFI and QSNR. By progressively including higher excited states, the thermometric precision is improved, highlighting the importance of the multilevel nature of the device.
We have also compared the transmon with its  harmonic oscillator approximation, which is often used as a simplified model for bosonic modes. In the temperature range considered in this work, the harmonic probe   generally yields a smaller   QFI with respect to its anharmonic counterpart, except for higher temperatures, while its QSNR grows monotonically within the considered temperature range, suggesting ever-improving sensitivity at higher temperatures. 
In contrast, the   multilevel Duffing model exhibits a larger maximum in the QFI at a finite optimal temperature. This behavior arises from the non-equidistant energy level structure of the transmon, which  fundamentally modifies the thermal response of the probe. These results demonstrate that neither the harmonic nor the two-level approximation provides a fully reliable description of transmon-based thermometry and underline the importance of accounting for the multilevel anharmonic spectrum when assessing temperature-estimation performance.

Second, we have quantified the role of the Josephson energy $E_J$ as a practical tuning parameter. While the charging energy $E_C$ is essentially fixed by fabrication,  $E_J$ can be adjusted in situ in flux-tunable transmons. Our analysis suggests that increasing $E_J$ slightly raises the optimal estimated temperature while reducing the corresponding maximaum of the QFI. This provides a  practical means of tailoring the thermometric response of the device.  In particular, by tuning  $E_J$, the maximum of the QFI can be shifted towards different temperatures, thereby optimizing the probe for local temperature estimation.

Then, we have addressed a theoretical issue that is often overlooked in the literature. The standard Duffing truncation  yields a quartic potential that is unbounded from below. Two physically consistent descriptions, based on confined quartic and sextic potentials, provide nearly identical QFI values in the thermometrically relevant regime, differing by only a few percent. By contrast, the Duffing approximation predicts a larger peak QFI, highlighting the role of higher-order corrections in quantitative thermometry.

Finally, we have explored the dependence of thermometric performance on the frequency $\omega$. Its variation shifts the entire energy spectrum and consequently modifies the temperature scale at which thermal excitations become relevant. We showed that  the optimal temperature  increases with $\omega$ whereas the corresponding  optimal QFI decreases. Therefore, probes designed to operate at higher temperatures exhibit a reduced peak thermometric sensitivity, highlighting a tradeoff between temperature range and estimation precision.

From a broader perspective, our work establishes the multi-level transmon as a versatile thermometer for nanoscale environments. Unlike standard thermometers that require separate calibration against a known reference, the transmon's well-characterized Hamiltonian, with $E_C$and $E_J$  known from independent measurements, allows the temperature to be inferred directly from steady-state population measurements, as demonstrated in recent experiments \cite{ThermoSC_exp}. By providing the ultimate precision limits through the QFI, our results offer a benchmark for current experimental thermometry schemes. Any practical measurement protocol (e.g., dispersive readout, or sequential $\pi$-pulse sequences) yields a classical Fisher information that is at most equal to the QFI  computed in this work.
In conclusion, we have demonstrated that the anharmonic energy-level structure of a superconducting transmon, often considered solely as a means to isolate a qubit, plays a decisive role in determining its thermometric performance. By moving beyond both the harmonic and two-level approximations, we have identified the optimal operating conditions and  addressed the pathologies coming from the quartic truncation. Our work provides both a guideline for the implementation  of transmon-based nanoscale thermometers and contribute to a deeper  understanding of the role of anharmonicity in  quantum-limited temperature estimation and pave the way to multipurpose superconducting devices. 

Our results also point to a natural direction for future investigations. While the present analysis focuses on the influence of the transmon spectrum, the thermometric performance of realistic devices is expected to depend additionally on material properties and fabrication-dependent features that affect the effective device Hamiltonian.
In realistic superconducting quantum circuits,  noise sources and fabrication imperfections and disturbance can alter the performances of the quantum thermometer. 
Effects such as 
dielectric loss, interface defects, spurious two-level systems, quasiparticle poisoning, flux noise, and junction inhomogeneity,  can  modify relaxation dynamics and 
broaden spectral lines, thereby affecting  the accuracy with which thermal populations are reconstructed. 
These effects do not alter the conceptual validity of the QFI benchmark, 
 they ultimately determine how closely an experimental protocol can approach it.
In this sense, the present results should be regarded as a first step toward a more complete characterization of transmon-based thermometry. 
Experimental investigations would be especially valuable for determining how the idealized thermometric performance predicted by our models is modified in realistic devices. Such studies could help distinguish the contributions of the intrinsic multilevel transmon spectrum from those arising from fabrication-dependent imperfections, material defects, and environmental noise.

More broadly, the present work suggests that materials optimization and thermal sensing should be treated as coupled design problems. 
A transmon fabricated in a low-loss environment with well-characterized junctions and reduced defect density is expected not only to perform better as a quantum processor, 
but also to provide a more reliable and eventually self-calibrating nanoscale thermometer for superconducting based quantum technologies. 
In this sense, advances in superconducting-device materials may be viewed as enabling conditions 
for the practical realization of quantum-limited thermometry in realistic circuits.

\acknowledgements
AM acknowledges the IRA Programme, project no. FENG.02.01-IP.05-0006/23, financed by
the FENG program 2021-2027, Priority FENG.02, Measure FENG.02.01., with the support of
the FNP. CB acknowledges support from MUR
 via the PRIN 2022  Project EQWALITY (Contract No. 202224BTFZ) and Piano di Sviluppo UniMi 2025. MGAP thanks Simone Cavazzoni for discussions. 
\bibliography{QFI_anharm}
\end{document}